\documentclass{elsart}

\usepackage[english]{babel}
\usepackage{doublespace}
\usepackage{amsmath}
\usepackage{amssymb}
\usepackage{epsfig}
\usepackage{graphicx}

\newcommand {\ignore}[1]{}

\def\lsim{\mathrel{\rlap{\lower4pt\hbox{\hskip1pt$\sim$}}
    \raise1pt\hbox{$<$}}}         
\def\gsim{\mathrel{\rlap{\lower4pt\hbox{\hskip1pt$\sim$}}
    \raise1pt\hbox{$>$}}}         

\makeatletter
\renewcommand{\fnum@table}{\textbf{\tablename~\thetable}}
\renewcommand{\fnum@figure}{\textbf{\figurename~\thefigure}}
\makeatother

\newcounter{myenumi}

\renewcommand{\themyenumi}{\roman{myenumi}}
{\end{list}}

\newlength{\myem}
\newcommand{\sep}[1]{#1}
\newcounter{mysubequation}[equation]

\newcommand{\bea}{\begin{eqnarray*}}
\newcommand{\eea}{\end{eqnarray*}}

\newcommand{\deltacp}{\delta_\mathrm{CP}}

\newcommand{\SM}{SU(3)$\times$\protect 
        \linebreak[0]SU(2)$\times$\protect\linebreak[0]U(1) }

\newcommand{\dm}[1]{{\Delta m^2_{#1}}}

\newcommand{\emt}{\epsilon_{\mu\tau}}

\begin{document}

\raisebox{8mm}[0pt][0pt]{\hspace{12cm}\vbox{IFIC/01-46\\MPI-Pht/2001-30\\
TUM--HEP--429/01}}

\begin{frontmatter}


\title{Non-standard Interactions:  Atmospheric versus Neutrino Factory Experiments}

\author{P.~Huber~$^{1,2,3}$} and 
\ead{Patrick.Huber@ph.tum.de}
\author{ J.~W.~F. Valle~$^{1}$}
\ead{valle@ific.uv.es}

\renewcommand{\thefootnote}{\alph{footnote}}

\address{$^{1}$ Instituto de F\'{\i}sica Corpuscular -- C.S.I.C., 
  Universitat de Val{\`e}ncia \\
  Edificio Institutos, Aptdo.\ 22085, E--46071 Val{\`e}ncia, Spain}

\address{$^{2}$
       Theoretische Physik, Physik Department, 
       Technische Universit{\"a}t M{\"u}nchen,\\
       James--Franck--Strasse, D--85748 Garching, Germany}
\address{$^{3}$
       Max-Planck-Institut f{\"u}r Physik, Postfach 401212, 
       D--80805 M{\"u}nchen, Germany }

\begin{abstract}
  We consider the potential of a generic neutrino factory (NUFACT) in
  probing non--standard neutrino--matter interactions (NSI).  We find
  that the sensitivity to flavour-changing (FC) NSI can be
  substantially improved with respect to present atmospheric neutrino
  data, especially at energies higher than approximately 50 GeV, where
  the effect of the tau mass is small. For example, a 100 GeV NUFACT
  can probe FC neutrino interactions at the level of few $|\varepsilon| < {\rm
    few} \times 10^{-4}$ at 99 \% C.L.
\end{abstract}

\end{frontmatter}


\section{Introduction}
\label{sec:introduction}

A long baseline neutrino
factory~\cite{Quigg:2000ap,Albright:2000xi,Autin:1998ga} offers a
unique tool for addressing basic questions in weak interaction and
flavour physics. One outstanding example is the quest for neutrino
mass and oscillations, which touches fundamental issues related to
Grand Unified Theories.  Motivated by the great discoveries of
underground experiments~\cite{SK-atm,atm-exp,SKconf,MACRO,sun-exp}
neutrino mass and oscillation searches have become the center of
attention in particle physics research. Apart from being motivated on
basic theoretical grounds~\cite{GRS,Schechter:1980gr,LR,Valle:1991pk}
neutrino masses and oscillations offer the simplest and most obvious
way to account for the observed
anomalies~\cite{Gonzalez-Garcia:2001sq}.  Nevertheless other
mechanisms, based on flavour changing non-standard neutrino
interactions have been suggested in connection with both
solar~\cite{Valle:1987gv,MSW,NSIrecent} and atmospheric
anomalies~\cite{Gonzalez-Garcia:1999hj,Fornengo:2000zp,otherFCfits,Val}
as well as other astrophysics
applications~\cite{Nunokawa:1996tg,Grasso:1998tt}. They can either
provide alternative solutions~\cite{Bergmann:2000gp} or else be
severely tested by the data, in the atmospheric case~\cite{Val}.  They
may arise in a number of theories beyond the Standard
Model~\cite{FCSU5,Barbieri:1995tw,Ross:1985yg,Fukugita:1988qe}, in
particular, in most (but not all) models of neutrino
masses~\cite{Schechter:1980gr}.

Using neutrinos from an accelerator in order to obtain an independent
confirmation of the non-accelerator physics results, is of fundamental
interest, as it will bring more light upon the issue of neutrino
masses and oscillations. This has been the focus of a number of
dedicated recent NUFACT
studies~\cite{Quigg:2000ap,Albright:2000xi,Autin:1998ga,Geer:1998iz}.
Following the recent suggestion in~\cite{Gago:2001xg} we propose the
use of a generic neutrino factory (NUFACT) to probe non--standard
neutrino--matter interactions (NSI). We show how indeed such an ideal
NUFACT can improve our present knowledge of non-standard FC neutrino
interactions well beyond what is presently attainable on the basis of
the latest atmospheric results and discuss the corresponding energy,
luminosity, energy resolution and tau detection requirements. We find
that, for example, a 100 GeV NUFACT can probe FC neutrino interactions
at the level of few $|\varepsilon| < {\rm few} \time 10^{-4}$ at 99 \% C.L.
without any assumption about tau charge identification. In contrast no
improvement is expected on non--universal (NU) neutrino interactions
beyond the present achieved sensitivity.

In order to compare the NUFACT sensitivities to NSI with present
atmospheric sensitivities, we will adopt the same approximation as in
ref.~\cite{Val}, i.e. we neglect the possible NSI in the production
and detection process of neutrinos. It is well understood that NSI can
be probed in a near detector with high accuracy (see e.g. \cite{nsi}).
However, the event rates in a near detector depend quadratically on
the strength of the NSI, whereas exploiting the non-standard matter
effects we obtain a \texttt{linear dependence} of the rates in a far
detector.  A combined treatment of NSI in production, propagation and
detection would lead to intriguing interference effects and is beyond
the scope of this letter.

\section{Interplay of Neutrino Oscillations and non-standard Interactions}
\label{sec:formalism}

The Standard Model can be extended to add neutrino masses in a variety
of ways~\cite{Valle:1991pk}.  In any massive neutrino gauge theory the
charged current (CC) weak interaction is characterized by the lepton
mixing matrix $K_{\alpha j}$.  This neutrino mixing matrix arises from the
unitary matrix ($U$) diagonalizing the neutrino mass matrix and the
corresponding unitary matrix ($\Omega$) diagonalizing the left-handed
charged leptons ($K = \Omega U$) and can be written in the following
parameterization~\cite{Schechter:1980gr}
\begin{equation} 
K = 
\begin{pmatrix} 
  c_{12}c_{13} & s_{12}c_{13} & s_{13}e^{-i\deltacp} \\
  -s_{12}c_{23} -c_{12}s_{23}s_{13}e^{i\deltacp} & 
  c_{12}c_{23}
  -s_{12}s_{23}s_{13}e^{i\deltacp} & s_{23}c_{13} \\
  s_{12}s_{23} -c_{12}c_{23}s_{13}e^{i\deltacp} & 
  -c_{12}s_{23}
  -s_{12}c_{23}s_{13}e^{i\deltacp} & c_{23}c_{13}
\end{pmatrix}. 
\label{Ut1} 
\end{equation}
where we see explicitly the usual three neutrino mixing angles
$\theta_{12}, \: \theta_{23}, \: \theta_{13}$ and one CP phase
$\deltacp$. This is the analogous of the CP phase found in the quark
sector, as the other two Majorana phases were set to zero, since they
are not observed in standard
total-lepton-number-conserving~\footnote{They could be seen only in
  $\Delta L =2$ processes, such as discussed in
  \cite{Schechter:1981gk}.}  oscillations.

The above $3\times 3$ form applies if there are no \SM singlet leptons,
such as the simplest models where neutrino masses arise
radiatively~\cite{Zee:1980ai,Babu:1988ki}.  In seesaw type
schemes~\cite{GRS,LR} the matrix $K_{\alpha j}$ is rectangular and contains
in general many more parameters: twelve mixing angles and twelve CP
phases in the three generation seesaw scheme~\cite{Schechter:1980gr}.
We assume, however, that singlets are all super-heavy so that the $K_{\alpha
  j}$ matrix can be well approximated by a unitary $3\times 3$ matrix and
parameterized as eq.~(\ref{Ut1}). This is in fact in agreement with the
scale of neutrino mass indicated by present neutrino anomalies.

All present neutrino data~\footnote{Except for the LSND anomaly, which
  requires a light sterile neutrino.  For recent discussions
  see~\cite{Hirsch:2000xe,Maltoni:2001mt}} can be accounted for by
eq.~(\ref{Ut1}). The two mass splittings $\Delta {m_\odot^2} \equiv
\dm{12}$ and $\Delta {m_{ATM}^2 \equiv \dm{23} \approx \dm{13}}$ as
well as the three neutrino mixing angles are all determined by global
fits of neutrino data~\cite{Gonzalez-Garcia:2001sq} which indicate
that two of the angles are large, $\theta_{13}$ being small due mainly
to reactor results~\cite{Chooz}. The recent SNO CC
data~\cite{Ahmad:2001an} adds support for the so-called LMA solar
neutrino solution~\cite{SNOfits}, which previously came only from
detailed solar recoil electron spectra~\cite{Gonzalez-Garcia:2000aj}.
Moreover, LMA is also consistent with the the observed SN~1987A
neutrino signal~\cite{SN87sol}. Thus in what follows we will take the
parameters appropriate to this solution. However the details of the
solar neutrino oscillation parameters do not significantly affect our
results.

Many theories beyond the minimal \SM model also lead to non-standard
neutrino interactions.  These include most models of generating
neutrino masses, which are generically accompanied by NSI, such as the
simplest seesaw type schemes~\cite{Schechter:1980gr,Valle:1991pk},
super-gravity SO(10) unified theories~\cite{Barbieri:1995tw}, models of
low energy super-symmetry with broken R parity~\cite{Ross:1985yg} as
well as some radiative models of neutrino mass~\cite{Fukugita:1988qe}.
Exceptional examples exist of situations where FC interactions are
unaccompanied by neutrino masses. Models involve neutral heavy leptons
at weak scale~\cite{MV,NHL} and some super-gravity SU(5) models
\cite{FCSU5}.  Such non--standard
interactions~\cite{Schechter:1980gr,Valle:1987gv,MSW,NSIrecent} can be
either flavour--changing (FC) or non--universal (NU).

  In Refs.~\cite{Gonzalez-Garcia:1999hj,Fornengo:2000zp,otherFCfits}
  the atmospheric neutrino data have been analyzed in terms of a pure
  $\nu_\mu \to \nu_\tau$ conversion in matter due to NSI.  The
  disappearance of $\nu_\mu$ from the atmospheric neutrino flux is due
  to interactions with matter which change the flavour of neutrinos. A
  complete analysis of the 79 kton-yr Super--Kamiokande data,
  including both the low--energy contained events as well as the
  higher energy stopping and through--going muon events from
  Super--Kamiokande and MACRO was given in Ref.~\cite{Val}.

  We therefore study an extended mechanism of neutrino propagation
  which combines both oscillation (OSC) and non--standard
  neutrino--matter interactions (NSI). In order to discuss the
  sensitivity of NUFACT to non-standard neutrino interactions we adopt
  the general Hamiltonian given by
\begin{eqnarray}
\label{eq:Hamilton}
\hat{H}=
K\,\left(
\begin{array}{ccc}0 & 0 & 0\\ 
0 & \Delta_{21} & 0\\
0 & 0 & \Delta_{31}
\end{array}\right)\,K^\dagger+
\left(\begin{array}{ccc}
V_e(r) & 0 & 0\\
0 & 0 & \epsilon_{\mu\tau}^f V_f(r)\\ 
0 & \epsilon_{\mu\tau}^f V_f(r) & \epsilon_{\mu\tau}^{'f} V_f(r)
\end{array}\right)\,.
 \end{eqnarray}
 containing both non--universal (NU) and flavour--changing (FC)
 interactions characterized by diagonal and off-diagonal entries in
 eq.~(\ref{eq:Hamilton}).  
 
 Note that, as usual, the matter potentials for neutrinos and
 anti-neutrinos differ in sign. In contrast, we assume that the new
 interactions are CP conserving. As a result the epsilon values have
 the same sign for neutrinos and anti-neutrinos. In the most general
 case the non-standard interactions might violate CP and the resulting
 phases could therefore affect the evolution. We will not consider
 this more complicated case in the following discussion.
 
 In addition to the five standard parameters (three angles and two
 mass splittings, if CP conservation is assumed) which describe the
 oscillation among three neutrinos there are, in the present scheme,
 also the $\epsilon_{\alpha\beta}$ and $\epsilon_{\alpha\beta}'$
 parameters characterizing the NSI of the neutrinos. Of the three
 possible channels, we choose to analyze in detail here only the
 $\nu_\mu - \nu_\tau$ transitions closely related to the atmospheric
 anomaly\footnote{For this reason we have neglected the
   $\epsilon_{\alpha\beta}$ and $\epsilon_{\alpha\beta}'$ involving
   the first generation in Eq.~(\ref{eq:Hamilton})}. The others will
 be discussed elsewhere.
 
 The relative importance of masses and NSI in the propagation of
 neutrinos is difficult to predict from basic principles and it is
 rather model-dependent. From a phenomenological point of view,
 however, atmospheric data imply that NSI can only play a sub-leading
 role~\cite{Val} in $\nu_\mu - \nu_\tau$ transitions.
 
 In order to gain some insight in the interplay between oscillation
 and NSI it is useful to reduce the problem to a two neutrino case by
 taking the limit $\dm{12}\to 0$.  In this case the rotation in the
 12-subspace also drops out, and therefore also the CP-phase becomes
 irrelevant~\cite{Schechter:1980gr}. This approximation is quite
 accurate for the $\nu_\mu - \nu_\tau$ transition at the energies and baselines
 considered for a neutrino factory experiment. In this limit only five
 parameters remain: three OSC parameters $\theta_{23}$, $\theta_{13}$ and
 $\dm{31}$ and our two NSI parameters $\epsilon_{\mu\tau}$ and $\epsilon'_{\mu\tau}$.
 Neglecting $\theta_{13}$ the transition probability is given
 by~\cite{Gago:2001xg}:
\begin{eqnarray}
  \label{eq:oszprob}
  P(\nu_\mu\to\nu_\tau)&=&\frac{B^2}{B^2+C^2}\sin^2\left(
\frac{L}{2}\sqrt{B^2+C^2}\right)\,,\\
\Delta_{13}&=&\frac{\dm{31}}{2E}\,,\nonumber\\
B&=&\Delta_{13}\sin 2\theta_{23} +2 \epsilon_{\mu\tau}V_f\,,\\
C&=&\Delta_{13}\cos 2\theta_{23}+\epsilon'_{\mu\tau}V_f\,.
\end{eqnarray}

\section{Simulating neutrino factory long baseline experiments}

In testing the effect of non-standard interactions in the $\nu_\mu - \nu_\tau$
transition it is essential to have a detector which is able to
identify $\nu_{\tau}$ events with a high efficiency. We are fully aware
that this is a very difficult goal to achieve in the design of a large
detector ($m\simeq 10\,\mathrm{kt}$). However, this work is partially
intended to show the benefits of such a detector in probing physics
beyond the standard model.  We will assume a detector with a mass of
$10\,\mathrm{kt}$ which is able to detect and identify $\nu_{\tau}$
interactions above a threshold of $4\,\mathrm{GeV}$ with a constant
efficiency of $\eta = 0.33$.  Basically there are the following
observables:

\begin{eqnarray}
  \mu^-\to\left\{\begin{array}{lr}\bar{\nu}_e\to\bar{\nu}_\tau&n^-(\bar{\nu}_\tau)\\ 
\nu_\mu\to\nu_\tau&n^-(\nu_\tau)\end{array}\right.\,,\quad
\mu^+\to\left\{\begin{array}{lr}\nu_e\to\nu_\tau&n^+(\nu_\tau)\\ 
\bar{\nu}_\mu\to\bar{\nu}_\tau&n^+(\bar{\nu}_\tau)\end{array}\right.\,.
\end{eqnarray}
As will be clear in section \ref{sec:results} the ability to identify
the charge of the tau is \emph{not} necessary for this particular
transition and therefore not assumed.  This leaves us with only two
observables:
\begin{equation}
 n^-:= n^-(\bar{\nu}_\tau)+n^-(\nu_\tau)\,,\quad n^+:=n^+(\nu_\tau)+n^+(\bar{\nu}_\tau)\,.
\end{equation}
where $ n^-$ and $ n^+$ denote the event numbers arising from the
neutrino factory operating in the two polarities.

In calculating the event rate spectra in a neutrino factory experiment
and for the treatment of the matter profile we follow the description
given in ref.~\cite{FHL}.  For the $\nu_\tau$ appearance channel we use the
cross-section given in~\cite{Paschos:2001np}, the $\bar{\nu}_\tau$ cross-section is 
assumed to be one half of this. We will show that
neglecting the tau mass~\cite{Gago:2001xg} is not a good approximation
especially for neutrino energies below $20\,\mathrm{GeV}$.  We also
take the energy resolution of the detector into account by modeling
it as a Gaussian, as described in ref.~\cite{FHL2001}.  The neutrino
factory delivers $2\cdot10^{20}$ useful muon decays of each polarity per
year for a period of 5 years. The energy of the neutrino factory is
indicated in each figure since it plays a crucial role in probing
non-standard interactions.

We now describe the Statistical Method we employ.  In order to
estimate the sensitivity to new physics we adopt the following
definition of $\chi^2$~\cite{FHL2001}
\begin{equation}\label{eq:chi}
  \chi^2=2\left(n^+-n^+_{\mathrm{OSC}}\right)+2n^+_{\mathrm{OSC}} \ln\frac{n^+_{\mathrm{OSC}}}{n^+}
+2\left(n^--n^-_{\mathrm{OSC}}\right)+2n^-_{\mathrm{OSC}}\ln\frac{n^-_{\mathrm{OSC}}}{n^-} \,.
\end{equation}
where $n^\pm_{\mathrm{OSC}}$ stands for the event rates which are
expected in the absence of NSI. This is readily obtained by leaving
all parameters as in the calculation of $n^\pm$ except that $\epsilon_{\mu\tau}$ and
$\epsilon'_{\mu\tau}$ are set to zero. Thus $\chi^2$ has two degrees of freedom,
therefore a value of $9.2$ corresponds to 99\%\,CL\,. Considering only
total rates, the sum over the energy bins is performed before $\chi^2$ is
calculated, whilst for an energy spectrum this sum is performed after
calculating $\chi^2$ for each bin.

The above $\chi^2$ is suitable to investigate the possible sensitivity to
the new effects arising from non-standard interactions. In order to
get reliable sensitivity limits it would be necessary in general to
take into account possible parameter correlations and to evaluate the
$\nu_\tau$ -appearance together with $\nu_\mu$--disappearance and
$\nu_\mu$--appearance. However for the $\nu_\mu - \nu_\tau$ transition this
complication is less relevant to the extent the relevant parameters
$\sin^22\theta_{23}$ and $\dm{31}$ that could be correlated with the NSI
parameters are already well determined by present atmospheric data.
As a consequence our results are basically unaffected by taking into
account these correlations. We have in fact verified this by an
explicit statistical analysis similar to that in ref.~\cite{FHL2001}.
The situation would be different for the $\nu_e - \nu_\tau$ transition since
this mode is controlled by $\sin^22\theta_{13}$ which is subject to much
higher uncertainties. For this reason this mode will be discussed
elsewhere~\cite{inprep}.

\section{Results}\label{sec:results}

\begin{figure}[htb!]
  \begin{center}
    \includegraphics[width=\textwidth]{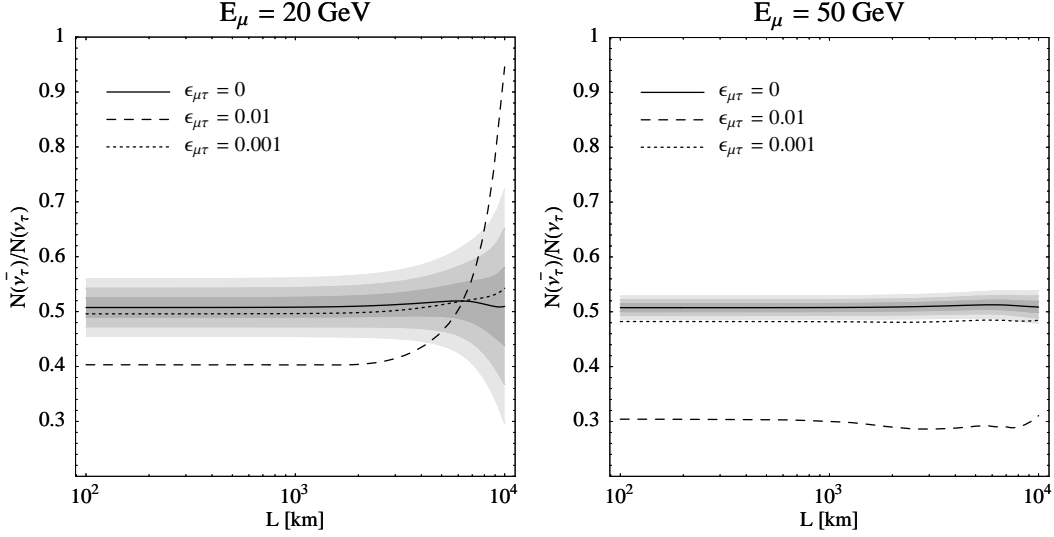}
    \caption{
      These figures show the ratio of observed $\bar{\nu}_{\tau}$ events
      from $\bar{\nu}_{\mu}$ to the observed $\nu_\tau$ events from $\nu_\mu$
      as a function of the baseline. The grey shaded bands indicate
      the Gaussian $1\sigma$, $2\sigma$ and $3\sigma$ statistical error on this
      ratio. The black solid line indicates the OSC prediction whereas
      the dashed lines indicate the deviation from this for different
      values of the FC parameter $\epsilon_{\mu\tau}$. The other parameters are
      $\sin^22\theta_{12}=0.78$, $\dm{21}=3.3\cdot 10^{-5}\,\mathrm{eV}^2$,
      $\sin^22\theta_{23}=0.97$, $\dm{31}=3.1\cdot 10^{-3}\,\mathrm{eV}^2$,
      $\sin^22\theta_{13}=0.02$, $\delta_{CP}=0$ and $\epsilon'_{\mu\tau}=0$.}
    \label{fig:ratio}
  \end{center}
\end{figure}

In order to highlight the effect of the non-standard interactions, we
show in figure~\ref{fig:ratio} the change in the ratio of $\bar{\nu}_\tau$
events (which arise from $\bar{\nu}_\mu$) to $\nu_\tau$ events (which arise from
$\nu_\mu$) for different values of the FC parameter $\epsilon_{\mu\tau}$.  If no NSI
interactions are present this ratio is basically constant $0.5$ (black
solid lines) since this transition is only very weakly influenced by
ordinary matter effects due coherent forward scattering off the
electrons~\cite{MSW}. The value of $0.5$ simply reflects the ratio of
the cross-section for $\bar{\nu}$ and $\nu$.  The grey shaded bands show
the Gaussian $1\sigma$, $2\sigma$ and $3\sigma$ errors on the standard ratio in the
absence of NSI neutrino interactions. The dashed lines indicate the
deviation from this for different values of the FC parameter $\epsilon_{\mu\tau}$.
One sees that for $\epsilon_{\mu\tau}$ at the per cent level the difference is
rather significant as long as the baseline is shorter than about 7000
km.  For the left hand panel with an muon energy of $20\,\mathrm{GeV}$
these errors increase drastically with the baseline. This is due to
the geometrical $L^{-2}$ loss of flux at large distances.  Comparing
the two panels one can easily see the importance of high energies in
order to obtain optimal sensitivity to the new physics. We have fixed
the OSC parameters as follows: $\sin^22\theta_{12}=0.78$ and $\dm{21}=3.3\cdot
10^{-5}\,\mathrm{eV}^2$ (suitable to account for the LMA solution of
the solar neutrino anomaly), $\sin^22\theta_{23}=0.97$ and $\dm{31}=3.1\cdot
10^{-3}\,\mathrm{eV}^2$ (suitable to account for the atmospheric
anomaly) and $\sin^22\theta_{13}=0.02$ in agreement with reactor results.
Finally we have assumed CP conservation, $\delta_{CP}=0$ and also exact
universality, $\epsilon'_{\mu\tau}=0$.
The measurement of this ratio would require
charge identification of the tau. It is shown here only for
illustrative purposes.

Note that the ratio in Fig.~\ref{fig:ratio} contains only one part of
the information contained in the event rates. For this reason we will use
the $\chi^2$ as defined in equation~\ref{eq:chi} in order to calculate
the sensitivity bounds to non-standard interactions.

\begin{figure}[htb!]
  \begin{center}
\includegraphics[width=1 \textwidth]{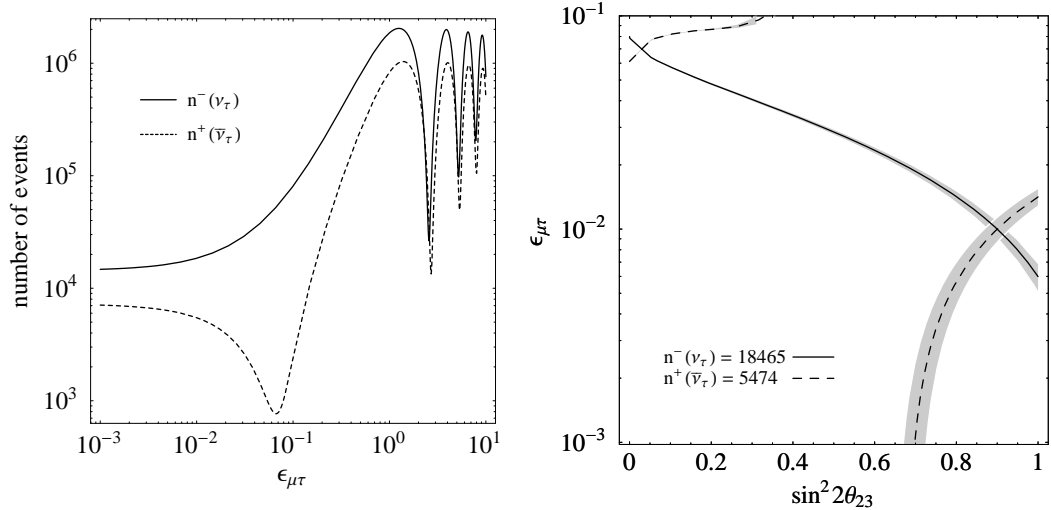}
\caption{
  In the right hand panel lines of constant event rates of the
  $\nu_\mu\to\nu_\tau$ transition (solid line) and of the $\bar{\nu}_\mu\to\bar{\nu}_\tau$
  transition (dashed lines) are shown in the $\sin^22\theta_{23}$ -
  $\epsilon_{\mu\tau}$ plane. The grey shaded band indicate the Gaussian $3\sigma$
  errors on these numbers.  The left hand panel shows a section
  across the right hand figure at $\sin^22\theta_{23}=0.9$. The baseline
  for both plots is $732\,\mathrm{km}$ and the muon energy is
  $50\,\mathrm{GeV}$. All other parameters are kept fixed to:
  $\sin^22\theta_{13}=0.02$, $\sin^22\theta_{12}=0.78$, $\delta_{CP}=0$,
  $\dm{31}=3.1\cdot 10^{-3}\,\mathrm{eV}^2$, $\dm{21}=3.3\cdot
  10^{-5}\,\mathrm{eV}^2$ and $\epsilon'_{\mu\tau}=0$.}
     \label{fig:ambi}
  \end{center}
\end{figure}

Before we do that let us highlight the important role played by
present atmospheric data by presenting Fig.~\ref{fig:ambi}.  For the
coming plots we use a baseline of $732\,\mathrm{km}$ and a muon energy
of $50\,\mathrm{GeV}$.  All other parameters are kept fixed as
previously to: $\sin^22\theta_{13}=0.02$, $\sin^22\theta_{12}=0.78$, $\delta_{CP}=0$,
$\dm{31}=3.1\cdot 10^{-3}\,\mathrm{eV}^2$, $\dm{21}=3.3\cdot
10^{-5}\,\mathrm{eV}^2$ and $\epsilon'_{\mu\tau}=0$.  In the left hand panel the
dependence of the event rates for the $\nu_\mu -\nu_\tau$ transition (solid
line) and for the $\bar{\nu}_\mu -\bar{\nu}_\tau$ transition (dashed line) on
the FC parameter $\epsilon_{\mu\tau}$ is shown for a fixed value of
$\sin^22\theta_{23}=0.9$.  For very small $\emt$ values the event rates in
both channels are nearly independent of $\emt$ and their ratio simply
reflects the ratio of the cross sections. For increasing values of
$\emt$ we now see that neutrinos and anti-neutrinos behave in a
opposite way.  This due to the fact that $V_f$ has a different sign
for neutrinos and anti-neutrinos. This behavior is also what would be
expected from a linearized version of eq.~\ref{eq:oszprob} as given in
ref.~\cite{Gago:2001xg}. At $\emt\simeq0.01$ however this simple picture
breaks down, the non-linearities of eq.~\ref{eq:oszprob} become very
important. Note that the transition probability only depends on $B^2$.
If the two terms contributing to $B$ become of the same order of
magnitude, i.,e.  $\Delta_{13}\sin 2\theta_{23}\simeq 2\emt V_f$ then the difference
between the sum of the two and their difference becomes maximal.
Therefore the ratio of the anti-neutrino rates to the neutrino rates
becomes minimal. Increasing $\emt$ further makes the difference
between neutrinos and anti-neutrinos smaller again. For large values
of $\emt$ the oscillation term $\Delta_{13}\sin 2\theta_{23}$ becomes negligible
and $\emt$ plays both the role of a mixing angle and provides the
leading contribution to the mass spliting. This leads to a strong
oscillating behavior in the event rates and also in the ratio, since
there are slightly shifted zeroes of the oscillation term
$\sin^2(L/2\sqrt{B^2+C^2})$. Thus there are in principle many
degenerate points in this case as can be seen from the right hand
panel. Here lines of constant event rates for the $\nu_\mu -\nu_\tau$
transition (solid line) and for the $\bar{\nu}_\mu -\bar{\nu}_\tau$ transition
(dashed line) in the $\sin^22\theta_{23}$ - $\emt$ plane are shown. There
are two points were the dashed and solid lines cross.  These points
have exactly the same physical observables and are therefore not
distinguishable in an experiment which uses only the total event
rates. However the point in the upper left corner can be excluded by
using the information of atmospheric neutrinos that
$\sin^22\theta_{23}>0.8$ and $\emt<0.02$~\cite{Val}.  There are also many
more possible solutions for $\emt$ values larger than $0.1$.  In order
to improve the knowledge on $\emt$ it is therefore necessary to
include atmospheric data.

\begin{figure}[htb!]
  \begin{center}
    \includegraphics[width=1 \textwidth]{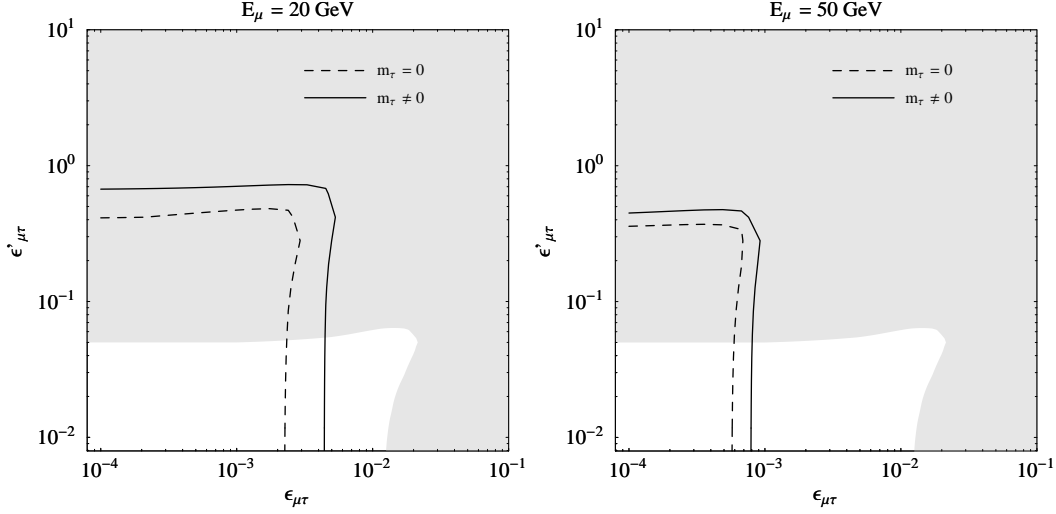}
    
\caption{
  Here the sensitivity limits in the $\epsilon_{\mu\tau}$- $\epsilon'_{\mu\tau}$ plane of a
  neutrino factory compared to atmospheric neutrino data are shown for
  an energy resolution of 50\%.  For comparison we also indicate with
  the grey shaded area the region presently excluded by the latest
  atmospheric data, taken from~\cite{Val}. All bounds are at 99\% CL.
  The dashed line is obtained by neglecting the tau mass and is only
  shown for comparison. The black lines are calculated with the
  correct cross-section. All other parameters are kept fixed to:
  values suitable to account for the present neutrino anomalies.
  Details in text.  }
    \label{fig:sens1}
  \end{center}
\end{figure}

\begin{figure}[htb!]
  \begin{center}
    \includegraphics[width=1 \textwidth]{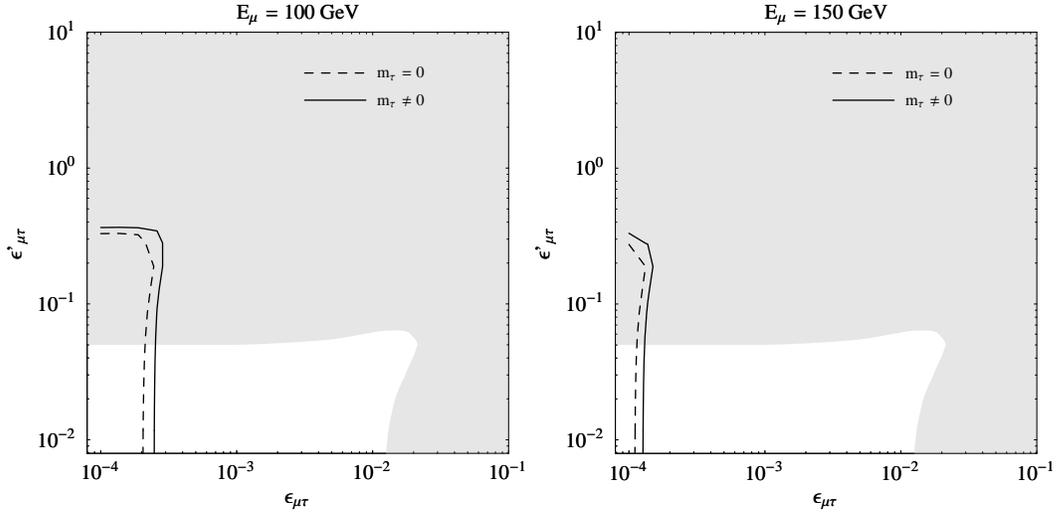}

\caption{
  Same as in Fig.~~\ref{fig:sens1} but for higher NUFACT energies.  }
    \label{fig:sens2}
  \end{center}
\end{figure}

We now come to our final results.  In figures~\ref{fig:sens1} and
~\ref{fig:sens2} we present our calculated NUFACT sensitivities to
non-standard neutrino interactions shown as black solid lines for
three different muon energies $20\,\mathrm{GeV}$, $50\,\mathrm{GeV}$,
$100\,\mathrm{GeV}$ and $150\,\mathrm{GeV}$. The baseline has been
chosen as $732\,\mathrm{km}$. The dashed lines show the bounds which
would be obtained by neglecting the tau mass
threshold~\cite{Gago:2001xg} in the cross-section. It is clearly
visible that especially for low energies this is not a good
approximation, since for example at 20 GeV one looses about 80~\% of
the events. For comparison we also indicate with the grey shaded area
the region presently excluded by the latest atmospheric data.  These
bounds are taken from~\cite{Val}.  The parameters were fixed as in
Fig.~\ref{fig:ratio}: $\sin^22\theta_{23}=0.97$, $\sin^22\theta_{13}=0.02$,
$\sin^22\theta_{12}=0.78$, $\delta_{CP}=0$, $\dm{31}=3.1\cdot
10^{-3}\,\mathrm{eV}^2$, and $\dm{21}=3.3\cdot 10^{-5}\,\mathrm{eV}^2$.
One sees that the limits on the FC parameter $\epsilon_{\mu\tau}$ can be improved
by approximately two orders of magnitude by a high energy neutrino
factory. For the low energy option (leftmost panel) the improvement in
the sensitivity at a neutrino factory is very small compared to the
present atmospheric bound. Note that the sensitivity on $\epsilon'_{\mu\tau}$
attainable at a long baseline neutrino factory is worse than the
present bounds by atmospheric data.

Note also that the above bounds \emph{do not} require tau charge
identification.  This is possible because the $\nu_e$-$\nu_\tau$
transition is suppressed by $\sin^22\theta_{13}$ (restricted to be
smaller than $0.1$ by the Chooz experiment~\cite{Chooz}) and ordinary
matter effects do not come into play at the distance of
$732\,\mathrm{km}$ considered here. In fact we have explicitly checked
that our results are unchanged if the signs of the NSI parameters get
reversed (in all possible combinations) with respect to what we have
assumed.  This approximation might break down at baselines longer than
1000 km or so.

\section{Discussion and Conclusions}

We have considered the potential of a long baseline neutrino factory
in probing non--standard neutrino--matter interactions.  We have found
that the sensitivity to flavour-changing NSI can be substantially
improved with respect to present atmospheric neutrino data, especially
at energies higher than 50 GeV or so, where the effect of the tau mass
is small. For example, a 100 GeV NUFACT can probe FC neutrino
interactions at the level of few $|\varepsilon| < {\rm few} \times 10^{-4}$ at 99 \%
C.L.  The analysis we have presented requires no tau charge
identification and is based only on total event numbers, with a modest
energy resolution at the 50~\% level. In order to be useful for more
refined studies a good detector energy resolution is required: for a
50\% energy resolution the results are basically the same as those
obtained when considering only total rates. It is doubtful whether a
better resolution can be achieved in practice for the channel
considered here because of hadronic tau decays. Finally note that the
quality of the atmospheric data plays a crucial role in setting this
limit by removing unwanted degeneracies in predicted event numbers.
In contrast the sensitivity on $\epsilon'_{\mu\tau}$ attainable at a long baseline
neutrino factory is worse than the present bounds by atmospheric data.
The role of a NUFACT in probing interactions is also complementary to
efforts to probe for similar flavour-changing effects in the charged
lepton sector and has the advantage of being totally model
independent.


\section*{Acknowledgments}

We would like to thank Michele Maltoni and Hiroshi Nunokawa for very
useful discussions.  This work was supported by Spanish grant
PB98-0693, by the European Commission RTN network HPRN-CT-2000-00148,
by the European Science Foundation network grant N.~86. PH was
supported by a fellowship of the European Commission Research Training
Site contract HPMT-2000-00124 of the host group.


\newpage

\bibliographystyle{phaip}

\end{document}